\documentstyle[prl,aps,floats,epsf]{revtex}
\font\msbm=msbm9  scaled \magstep 1    
\def\kappa{\hbox{\msbm\char"7B}}       
\begin{document}
\draft
\tighten
\twocolumn[\hsize\textwidth\columnwidth\hsize\csname
@twocolumnfalse\endcsname
\title{ Paramagnetic Meissner effect in superconductors \\
from self-consistent solution of Ginzburg--Landau equations}
\author{G.F. Zharkov}
\address{P.N. Lebedev Physics Institute, Russian Academy of Sciences,
 Moscow, 117924, Russia}
\date{July, 2000}
\maketitle
\begin{abstract}
The paramagnetic Meissner effect (PME) is observed in small superconducting
samples, and a number of controversial explanations of this effect are
proposed, but there is as yet no clear understanding of its nature. In the
present paper PME is considered on the base of the Ginzburg-Landau theory (GL).
The one-dimensional solutions are obtained in a model case of a long
superconducting cylinder for different cylinder radii $R$, the GL-parameters
$\kappa$ and vorticities $m$. Acording to GL-theory, PME is caused by the
presence of vortices inside the sample. The superconducting current flows around
the vortex to screeen the vortex internal field from the bulk of the sample.
Another current flows at the boundary to screen the external field $H$ from
entering the sample. These screening currents flow in opposite directions and
contribute with opposite signs to the total magnetic moment (or magnetization)
of the sample. Depending on $H$, the total magnetization $M$ may be either
negative (diamagnetism), or positive (paramagnetism). The detailed study of a
very complicated saw-like dependence $M(H)$ (and of other characteristics),
which follow from the self-consistent solutions of the GL-equations, is
presented.

\end{abstract}
\pacs{ }
\vskip 2pc ] 
\narrowtext

\section{INTRODUCTION}
A superconductor, placed in magnetic field and cooled down through the
transition temperature, expels the magnetic flux. This phenomen, known as the
diamagnetic Meissner effect (DME), is one of the most essential properties of
superconductors. Suprisingly, several recent experiments have shown that some
superconducting samples may attract magnetic field, which corresponds to the
so-called paramagneic Meissner effect (PME). Originally, PME was observed in
small size high-temperature superconductors (HTS), what has prompted the
appearance of a number of theories, attributing this effect to a
non-conventional superconductivity in these materials. (Mention such
explanations, as spontaneous currents due to $\pi$-contacts, vortex pair
fluctuations, $d$-wave superconductivity, orbital glass, the Josephson
junctions, the Andreev reflection of pairs at the boundary, and others.
The numerous references to the experimental and theoretical work on PME may be
found in recent papers [1--7].)\, Later, the anomalous PME was observed also
in conventional low-$T_c$ supercontuctors and was considered as intrinsic
property of any finite-size superconductor.

In [2--5] the PME was studied in a case of conventional superconductors, using
the Ginzburg--Landau (GL) system of equations. Here such conceptions were put
forward, as the flux capture and its compression [2], the hysteresis transitions
and the role of experimental setup [3], the metastability of the vortex
configuration due to the influence of the sample surface [4], the disbalance
of screening currents in metastable vortex states [5].

As was noticed in [5], the GL-theory treats PME in finite-size superconductors
as purely electrodynamic effect, caused by presence of vortices inside the
superconductor. If the vortex exists inside the sample, the superconducting
current $j_s$ flows around the vortex center in a direction to screen out the
vortex field from entering the bulk of the sample. If the sample is placed in
external magnetic field $H$, the additional surface current flows to screen out
the field $H$ from entering the sample. These two currents have opposite
directions and produce contributions of opposite sign to the sample magnetic
moment (or, magnetization) ${\bf M}=(1/2c)\int [{\bf j}_s\times {\bf r}]dv$.
The resulting value of $M$ (in the direction of $H_z>0$) may be either negative
(this corresponds to diamagnetic susceptibility), or positive (this corresponds
to paramagnetic susceptibility), depending on $H$ magnitude.

In the present paper the origination of PME within the GL-theory approach is
studied in more details, as compared with [5], including both type-I and
type-II superconductors, for various sample dimensions and different values of
GL-parameter $\kappa$. In Sec. 2 the GL-equations for the order parameter
$\psi$ and magnetic field $B$ inside the superconductor are written. In Sec. 3
the results of numerical calculations of non-linear system of equations are
presented by a number of graphics, accompanied by necessary comments. Sec. 4
contains a short resume and discussion of the results.

\section{EQUATIONS}

Below the case is considered of a long superconducting cylinder of radius $R$,
in the external magnetic field $H\ge 0$, which is parallel to the cylinder
element. In the cylindrical co-ordinates the system of GL-equations may be
written in dimensionless form [5]
$$
{ {d^2U} \over {d\rho^2} } - {1\over \rho}{ {dU}\over {d\rho} } -
\psi^2 U=0,                                                    \eqno(1)
$$
\vskip -0.5cm
$$ { {d^2 \psi} \over {d\rho^2} } + {1\over\rho}
{{d\psi}\over{d\rho}} + \kappa^2 (\psi - \psi^3) - { {U^2 } \over {\rho^2}
}\psi =0.                                                     \eqno(2)
$$
Here $U(\rho)$ is the dimensionless field potential; $b(\rho)$ is the
dimensionless magnetic field; $\psi(\rho)$ is the normalized order parameter;
$\rho=r/\lambda$, $\lambda$ is the field penetration length;
$\lambda=\kappa\xi$, where $\xi$ is the coherence length, $\kappa$ is the
GL-parameter. The dimensioned potential $A$, field $B$ and current $j_s$ are
related to the corresponding dimensionless quantities by the formulae [5]:
$$ A={ {\phi_0} \over {2\pi\lambda} }{ U + m \over \rho },\quad
B={ {\phi_0} \over {2\pi\lambda^2} }b,\quad
b={1\over \rho}{{dU}\over{d\rho}},
$$
\vskip -0.5cm
$$
j(\rho)=j_s\Big/ { {c\phi_0} \over {8\pi^2\lambda^3} }=
-\psi^2 {U\over \rho},\quad  \rho = {r\over \lambda}.              \eqno(3)
$$
(The field $B$ in (3) is normalized by $H_\lambda=\phi_0/(2\pi\lambda^2)$,
with $b=B/H_\lambda$; instead of $H_\lambda$ one can normalize by
$H_\xi=\phi_0/(2\pi\xi^2)$, or by
$H_{\kappa} = \phi_0/(2\pi\xi\lambda) = H_\xi/\kappa$. The coefficients in (1),
(2) would change accordingly.) The vorticity $m$ in (3) specifies how many flux
quanta are associated with the vortex, centered at the cylinder axis (the
giant-vortex state).

The boundary conditions to Eq. (1) are [5]:

$$ U\big|_{\rho =0} = -m,\quad
 \left. { {dU}\over{d\rho} }\right|_{\rho =\rho_1}=h_\lambda.      \eqno(4)
$$
where $\rho_1=R/\lambda$, $h_\lambda=H/H_\lambda$.

The boundary conditions to Eq. (2) are [5]:
$$
\left. {d\psi \over d\rho} \right|_{\rho =0} =0, \quad
\left. {d\psi \over d\rho} \right|_{\rho=\rho_1} =0 \quad (m=0),
$$
\vskip-0.9cm
$$
                        \eqno(5)
$$
\vskip-0.9cm
$$
\psi|_{\rho=0}=0,\quad
\left. { d\psi \over d\rho} \right|_{\rho=\rho_1} =0 \quad (m>0).
$$

The magnetic moment (or, magnetization) of the cylinder, related to the unity
volume, may be written in a form
$$
{M\over V}={1\over V}\int  {B-H \over 4\pi }dv = { B_{av}-H \over 4\pi },
$$
$$
B_{av}={1\over V}\int B({\bf r})dv={1\over S}\Phi_1,
$$
where $B_{av}$ is the mean field value inside the superconductor, $\Phi_1$ is
the total magnetic flux, confined in the cylinder. In the normalization (3),
denoting $\overline{b}=B_{av}/H_\lambda$, $h_\lambda=H/H_\lambda$,
$M_\lambda=M/H_\lambda$, one finds
\begin{eqnarray*}
\qquad\qquad\quad 4\pi M_\lambda=\overline{b}-h_\lambda, \quad
\overline{b}={2\over\rho_1^2}(U_1+m), \quad \quad   (6)  \\
\phi_1={\Phi_1 \over \phi_0}=U_1+m, \quad
U_1=U(\rho_1),\quad \rho_1={R\over\lambda}.
\end{eqnarray*}

Accordingly, the normalized Gibbs free energy of the system may be written as
[5]
$$
\Delta g=\Delta G\Big/ \left( { H_{c{\rm m}}^2 \over 8\pi } V \right)=
g_0-{8\pi M_\lambda \over \kappa^2} h_\lambda +{4m \over \kappa^2}
{b(0)-h_\lambda \over \rho_1^2},                          \eqno(7)
$$
\vskip -0.7cm
$$
g_0={2\over\rho_1^2} \int_0^{\rho_1}
\rho d\rho \left[ \psi^4-2\psi^2+{1\over\kappa^2} \left(
{d\psi \over d\rho} \right)^2 \right].
$$
Here $\Delta G=G_s-G_n$ is the difference of free energies in superconducting
and normal states; $b(0)=B(0)/H_\lambda$, $B(0)$ is the magnetic field at the
cylinder axis; $H_{c{\rm m}}=\phi_0/(2\pi\sqrt{2}\lambda\xi)$ is thermodinamical
critical field of massive superconductor; $g_0$ is the condensation energy with
account for the order parameter gradient. The expressions (6), (7) will be used
below for calculating the corresponding quantities.

\section{NUMERICAL RESULTS}

The system of equations (1), (2) was solved numerically by the method, described
in [5,8]. For every fixed set of parameters $R, H, m, \kappa$ the unique
self-consistent solution for $\psi(\rho)$ and $U(\rho)$, satisfying the boundary
conditions (4), (5), was found. If the external field $H$ exceeds some critical
value $H_c^{(m)}$ ($H>H_c^{(m)}$), no superconducting solution with $\psi\ne 0$
is possible. The phase diagrams, which separate the superconducting
($\psi\ne 0$) and normal ($\psi\equiv 0$) regions, are presented in Fig. 1 for
different vorticities $m$ and parameters $\kappa$.

The critical field of the vortex-free Meissner state ($m=0$) is distinguished
by thick curve. The superconducting region lays between this curve and the axis
$h_\xi=0$ ($h_\xi=H/H_\xi$). For $m>0$ the superconducting region is inside of
the corresponding $m$-curve (thin line). To every point of the superconducting
$m$-region corresponds some solution of Eqs. (1)--(5). The transformation of
solutions, when the superconducting regions are crossed along the lines
$R_\lambda\equiv R/\lambda={\rm const}$ (see the dotted horizontal lines), is
illustrated below by Figs. (2)--(4).

In Figs. 1($a,b$) the dotted vertical line $h_\xi=1.695$ represents the critical
field $H_{c3}=1.695H_{c2}$ ($H_{c2}=\phi/(2\pi\xi^2)\equiv H_\xi$), at which
the surface superconductivity [9,10] nucleates in large ($R_\lambda\gg 1$)
samples. It is seen, that in type-II superconductors ($\kappa>1/\sqrt{2}$) the
superconducting states ($m>0$) are possible, which lay to the right of the
line $m=0$. One may say, that these states are enhanced by the external field.
In contrast to this, in type-I superconductors (see Fig. 1($c$)) the states
$m>0$ lay to the left of the line $m=0$, so they are weakend by the field.

Fig. 2 illustrates the superconductor behavior ( $R_\lambda=2, \kappa=2$), when
the field $h_\xi$ varies. In Fig. 2($a$) the regions are depicted, where the
state $m$ can exist ($\psi_{max}$ is the maximal value of the order parameter
in the corresponding $m$-state). For instance, the state $m=4$ (which is
distinguished for clarity by thick line) can exist only in the interval of
fields, where $\psi_{max}>0$. The state $m=11$ can exist only in sufficiently
large fields, $h_\xi\sim 2$ (see also Fig. 1($a$)).

Fig. 2($b$) shows the field dependence of the cylinder magnetization in
different $m$-states. For instance, the magnetization ($-4\pi M_\lambda$) in
the state $m=4$ may be either positive (this corresponds to diamagnetic
susceptibility, $M_\lambda<0$), or negative (this corresponds to paramagnetic
susceptibility, $M_\lambda>0$), depending on the field value. The vortex-free
Meissner state ($m=0$) is diamagnetic for all fields.

Fig. 2($c$) shows the normalized free-energy difference, $\Delta g$, for the
states with different $m$. At the points, where the curves with neighboring
values of $m$ cross each other (marked by open circles), the equilibrium
transitions between the states $m$ and $m\pm 1$ became possible. These
equilibrium transitions are marked by the dashed vertical lines in Fig. 2($b$).
If the equilibrium transitions are only allowed, the magnetization curve would
always be diamagnetic ($-4\pi M_\lambda>0$), and have a reversible saw-like
shape. However, if the metastable states are also allowed, the magnetization
curve would look very different. This is illustrated by Figs. 3($a,b,c$) for
$\kappa=1$ and by Figs. 3($d,e,f$) for $\kappa=0.5$.

Consider Fig. 3($c$), the state $m=2$. That part of the curve $m=2$, which lays
to the left of the equilibrium point ($eq$), is metastable (because the lower
energy states $m=0$, or $m=1$, exist). If the field $h_\xi$ is diminished from
its equilibrium value (in the field-cooling regime), the state $m=2$ can persist
only down to the point $\alpha$, where the forced transition to other state
should occure. The nearest laying energy state is $m=1$, it corresponds to the
relative minimum of the free energy (the state $m=0$ corresponds to the absolute
minimum). If the transitions between the nearest laying energy minimae are
possible, then at the point $\alpha$ the transition from the state $m=2$ to the
state $m=1$ may occure. In Fg. 3($b$) this would mean the transition from the
paramagnetic state $m=2$ (which lays to the left of the dashed-dotted vertical
line, passing through the point $-4\pi M_\lambda=0$) to the paramagnetic state
$m=1$. If the field $h_\xi$ is now increased, the state $m=1$ may survive up to
the point $\delta$, where the forced transition from the diamagnetic state $m=1$
to the diamagnetic state $m=3$ may occure. At the point $\beta$ the transition
from paramagnetic part of the curve $m=3$ to the paramagnetic part of the curve
$m=2$ occures. At the point $\gamma$ the transition between diamagnetic ($m=0$)
and paramagnetic ($m=3$) states may happen. Note, that all paramagnetic states
are metastable (they lay to the left of the corresponding equilibrium points).
Evidently, the transitions between various states, shown in Figs. 3($a,b,c$),
may lead to the hysteresis behavior of $M(H)$ (see also [3,4]).

[Note, that for larger values of $R_\lambda$ and $\kappa$ the reversible jump
transitions to the edge-suppressed states [5] of the same vorticity $m$ are
also possible, which might additionally complicate the dependence $M(H)$.]

The analogous transitions between paramagnetic and diamagnetic states are also
possible in type-I superconductors (see Figs. 3($d,e,f$)). Notice, that the
curve $m=0$ terminates at the point $\gamma$, where $\Delta g>0$. This part of
the curve corresponds to the "superheated" superconducting state, considered
first by Ginzburg [11] for type-I superconductors in the approximation
$\psi={\rm const}$ (see also [12]). The point $\gamma$ corresponds to the
maximal superheating. The only possible transition here may be to the normal
state ($\psi=0$). If the field $h_\xi$ is now diminished, the normal state may
be "supercooled" [11,12] down to the point $\delta$ ($h_\xi=1.49$), where the
superconducting state recovers. The arrows in Fig. 3($d$) mark the boundary of
the hysteresis loop. (The supercooled states exist also in a case
$\kappa>1/\sqrt{2}$. They additionally complicate the picture and are not shown
in Figs. 3($a,b,c$).)

As is seen from Figs. 2,3, each magnetization curve $(-4\pi M_\lambda^{(m)})$
has the diamagnetic and paramagnetic parts. Consider points of maximal
diamagnetism and maximal paramagnetism. Fig. 4 illustrates the space
distributions $\psi(r)$, $b(r)$, $j(r)$ at these points for several values of
$m,\kappa,R_\lambda$. The curves {\it 1} (thick lines in Fig. 4) correspond to
maximal paramagnetism ($M_\lambda>0$), the curves {\it 3} (thin lines)
correspond to maximal diamagnetism ($M_\lambda<0$), the curves {\it 2} (dashed lines)
represent points, where $M_\lambda=0$.

Figs. 4($b,e,h$) show, that for paramagnetic curves {\it 1} the magnetic field
inside the sample is greater, than the applied field $h_\lambda$
($b(r)>h_\lambda$, $h_\lambda=\kappa^2 h_\xi$).
This means, that the influence of the internal vortex field $B(0)$ is greater,
than the influence of the external field $H$. The relative suppression of the
order parameter, seen for the curves {\it 1} in Figs. 4($a,d,g$), is caused
mainly by the internal field $B(0)$. The same is seen in Figs. 4($c,f,i$),
where all the currents {\it 1} are paramagnetic ($j>0$), they screen mainly
the internal vortex field.

For larger fields $H$ (the curves {\it 2}), the role of external magnetic field
is more pronounced. Now, partly $b(r)>h_\lambda$ (the paramagnetic behavior),
and partly $b(r)<h_\lambda$ (the diamagnetic behavior). Analogously behaves the
current $j$: part of the current is $j>0$ (paramagnetic screening of the vortex
field), and partly $j<0$ (diamagnetic screening of the external field). These
two parts of $j$ counterbalance each other, resulting in zero magnetization,
$M_\lambda=0$.

The order parameter $\psi(r)$ in the case {\it 2} is enhanced by a larger field
(compare the curves {\it 2} and {\it 1} for $\psi$). This reflects the
reentrant behavior of small-size samples: the superconductivity, which was
impossible in small external field, is possible again in larger fields (see
Figs. 2,3).

When the field $H$ is increased further (the curves {\it 3} in Fig. 4), the
sample passes to diamagnetic state ($M_\lambda<0$): the field $b(r)<h_\lambda$
everywhere, the current $j$ screens predominantly the external field ($j(r)<0$
near the surface), the order parameter $\psi(r)$ displays the suppression by
strong external field $H$.

In the foregoing discussion the case was considered, when the external field
$H$ has been varied, but the temperature $T$ was fixed (see the horizontal
dotted lines $T={\rm const}$ in Fig. 1). In some experiments (for instance, [7])
the field $H$ was fixed, but the temperature $T$ varied. In this case the
transitions between different diamagnetic and paramagnetic states (see
Figs. 3($b,c$)) are also possible.

Indeed, in the vicinity of $T_c$ the
parameters $\xi$ and $\lambda=\kappa\xi$ depend on the sample temperature $T$
by the law $\xi(T)=\xi_0 (1-T/T_c)^{-1/2}$. Using notations: $t=T/T_c$,
$R_\lambda=R/\lambda$, $h_\xi=H/H_\xi$, $H_\xi=\phi_0/(2\pi\xi^2)$, one has
two relations: $R_\lambda=r_0\sqrt{1-t}$ and $h_\xi=h_0/(1-t)$, or
$$
1-t={R_\lambda^2 \over r_0^2}={h_0\over h_\xi},               \eqno(8)
$$
where $r_0=R/\lambda_0$, $h_0=H/H_{\xi 0}$, $H_{\xi 0}=\phi_0/(2\pi\xi_0^2)$,
$\lambda_0=\kappa\xi_0$. From Eqs. (8) the dependence between $R_\lambda$ and
$h_\xi$ follows:
$$
R_\lambda={ A\over \sqrt{h_\xi} }, \quad
A={R\over \lambda_0}\left( { H \over H_{\xi 0} } \right)^{1/2}
\equiv {1\over\kappa}\left( {2\pi R^2 H \over \phi_0} \right)^{1/2}. \eqno(9)
$$
The dashed curves in Fig. 1($b$) are the lines, along which the representation
point $(R_\lambda, h_\xi)$ moves, when $H={\rm const}$ and the temperature $T$
varies only. Taking $\kappa=1$, $H=100$\,Oe, one has $A=1$ for
$R=1.8\cdot 10^{-5}$\,cm, and $A=3$ for $R=5.4\cdot 10^{-5}$\,cm. Notice the
essential difference in the dashed curves positions. For smaller $R$ ($A=1$)
the system remains mainly in the diamagnetic state $m=0$, so no PME is possible.
For larger $R$ ($A=3$) the dashed curve $H={\rm const}$ crosses several of
$m$-state phase boundaries. This more complicated situation is illustrated by
Fig. 5.

In Fig. 5 the temperature dependencies are shown: of the order parameter
$\psi_{max}^{(m)}$, of the magnetization ($-4\pi M_\lambda^{(m)}$), and the
Gibbs free energy $\Delta g^{(m)}$ in several $m$-states. It is seen, that the
first superconducting state ($\psi\ne 0$), encountered in the temperature-cooled
mode, is $m=2$ (see Fig. 1($b$), the point $\mu$). This is the state of the
surface superconductivity in a finite-dimension sample. The magnetization curve
is diamagnetic ($M_\lambda^{(2)}<0$), it starts from zero, as the second order
phase transition. When the temperature $t=T/T_c$ is reduced further, other
$m$-states became possible. If the temperature $t$ is swept up and down (as in
[7]), alongside with the equilibrium transitions between diamagnetic states (at
the points $\alpha$ and $\beta$ in Figs. 5($b,c$)) the transitions between
various metastable branches of the curves may also occure. For instance, at the
termination point $\gamma$ the jump transition from the diamagnetic state $m=0$
to the nearest energy laying paramagnetic state $m=4$ is possible; at the
termination point $\delta$ the jump from the paramagnetic state $m=4$ to the
diamagnetic state $m=1$ may occure. The up and down arrows at the points
$\gamma$ and $\delta$ in Fig. 5($b$) mark the possible hysteresis loop in the
$M(T)$-dependence (compare with the experimentally found dependencies [7]).

A more detailed comparison of the present calculations with experiments on PME
is not possible, because the model case of a long circular cylinder approximates
the real specimens geometry very crudely. However, it is instructive to see the
results of GL-theory, because they suggest some ideas, which should be
viewed with attention. For instance, in the scope of GL-theory the PME looks
more like a bulk effect, it is caused by the balance of currents, flowing inside
the sample. The role of the surface seems not to be crucial, though in the
finite-hight extended samples [7] the role of the boundary may be important. It
also looks probable, that PME might be observed in the proximity structures [6],
where the superconducting cylinders are embodied into the normal-metal
enviroment, or in a system of microholes inside the superconducting matrix [13].

Below are comments on some of the papers, where the PME was first considered
within the GL-approach.

In [15] the hypothesis was put forward of a normal layer, which might be formed
inside the superconducting plate under the influence of the external field $H$
and is situated at some distance from the plate boundary. According to [15],
the conception of such a layer might help in understanding the nature of PME in
superconductors. Note, that in the homogeneous GL-theory no such layer is
possible, though the basic idea of [15] is, probably, correct. Indeed, if a
chain of vortices is formed near the plate surface (with the order parameter
vanishing at their axes), the average vortices distribution may be visualized
as a normal layer, in accordance with [15].

In [2] the conception of the flux compression was proposed to explain the PME.
According to [2], the giant $m$-vortex state, which is formed at sufficiently
high field $H$, persist at smaller $H$ in a form of the flux-compressed state,
with the eventual onset of the paramagnetic response. Some evidence of the
possible PME in a small-radius cylinder was obtained in [2] by self-consistent
solution of GL-equations. Note, that our calculations are along the same lines
(see Figs. 2, 3 and also [5]), but they show, that the giant $m$-vortex state
can persist only in a finite field interval, outside which the vorticity $m$
changes. As it follows from Figs. 4($b,e,h$), the imposition of the external
field $H$ leads not to a field compression, but, better to say, to a field
submission effect. Indeed, at smaller $H$ (the curves {\it 1}) the external
field is effectively screened out ($b(r)>h_\lambda$); at larger $H$ (the curves
{\it 2}) the Meissner screening is not so effective and the external field is
partly submitted into the sample ($h_\lambda>b(r)$ near the boundary); still at
larger $H$ (the curves {\it 3}) the external field floods almost all the
interior ($h_\lambda>b(r)$ everywhere), with eventually total submission to the
external field ($b(r)=h_\lambda$ in the normal state). As was mentioned above,
the appearance of PME in this picture is due to the redistribution of the
screening currents, flowing inside the specimen.

[There exist some doubts concerning the self-consistency of the solutions,
found numerically in [2]. Thus, in Fig. 4($a$) [2] the order parameter must
depend linearly on $r$ for $r\ll 1$. The local induction $b(r)$ in Figs. 4($c$)
and 5($a$) [2] is maximal not at $r=0$ (where the vortex axis is situated), but
at some distance $r\ne 0$, which seems strange, etc. Most closely our
calculations correlate with those in [3,4,15], where many fragments of the
picture, presented above, may be found, though no concise formulation of a
possible nature of PME in conventional superconductors was given.]

\section{Conclusions and discussion}

The GL-theory of a long superconducting cylinder, which accounts for the
transitions between centrally symmetric $m$-states (the giant-vortex states),
gives only qualitative picture of the processes inside the real superconductor.
In reality, the giant-vortex may split itself into the asymmetric multy-vortex
configuration of the same vorticity $m$. The specimen finite hight in
$z$-direction, and the boundary conditions at the surface, might also be
important for calculating the real vortex structure. All this makes the
analysis more difficult. [The approximate GL-description of the multy-vortex
configurations in mesoscopic 3-dimensional samples may be found in [3,4].)\,
However, even in a general case of arbitrary vortex configuration, there will
be two currents, flowing in opposite directions: one - to screen out the
combined fields of the vortices, and second - to screen out the external field.
Thus, basing on the GL-model calculations, it seems reasonable to offer the
following qualitative scenario for the appearance of PME in finite-dimension
samples.

According to GL-theory, the paramagnetism in conventional low-$T_c$
superconductors may be caused by the presence of
vortices inside the superconductor. The vortices, which are formed inside the
sample in the field-cooling regime, may be pinned to the boundary, which is a
source of inhomogeneity. The directions of the external and of the vortex
magnetic fields are the same, but the directions of currents, which flow to
screen these fields from the bulk of the sample, are opposite (clock-wise and
counter clock-wise). The screening currents are responsible for two
contributions of opposite sign to the total magnetic moment (or, equivalently,
magnetization) of the sample: $M=M_{para}+M_{dia}$. The distributions of the
fields and currents inside the sample varies with $H$, so the measured
magnetization may be either positive ($M>0$, PME), or negative ($M<0$, DME), or
$M=0$, depending on $H$.

The pinning of the vortices to the boundary is a non-linear effect, described
by the system of GL-equations. Each of the vortex configurations of fixed
vorticity $m$ can exist only within a limited interval of fields $H$. When the
field takes the value outside this interval, the existing vortex configuration
reconstructs, and a new configuration with different vorticity $m'$ establishes
(see Fig. 3). The jump on the magnetization curve may accompany such
reconstruction. The new configuration may reveal either PME, or DME, depending
on the resulting state characteristics. If the system may occupy only the ground
state of minimal free energy, then no PME is possible, and the behavior $M(H)$
should be completely reversible. The presence of PME and of the hysteresis is a
signal of metastability. The examples of transitions between various metastable
states are presented in Fig. 3, both for type-I and type-II superconductors.

The proposed qualitative GL-scenario looks rather general and one may attempt
to apply it also to the case of HTS ($\kappa\gg 1$), where PME is also observed
in a field-cooling regime, is also characterized by the hysteresis, by the jump
transitions between PME and DME branches of magnetization, and by the
transitions between various quantum flux states (see, for instance, [14]).
Whether PME in HTS can be explained by the presence of vortices, or by more
exotic mechanisms, mentioned in the Introduction, is a question for further
experimental and theoretical study.

\section{Acknowledgments}

I am grateful to V.L. Ginzburg for the interest in this work. I also thank
H.J. Fink, who pointed out to the paper [16], where the edge-suppressed states
(discussed recently in [5]) were first encountered (see the footnote in [16]).

\centerline{\bf Figures captions}

Fig. 1. The superconducting phase boundaries of vortex states with different
vorticities $m$ (the numerals at the curves). ($a$) -- $\kappa=2$; ($b$) --
$\kappa=1$; ($c$) -- $\kappa=0.5$. The vertical dotted line $h_\xi=1$
corresponds to the critical field $H_{c2}$, the line $h_\xi=1.695$ corresponds
to the critical field $H_{c3}$. (Here $R_\lambda=R/\lambda$, $h_\xi=H/H_\xi$,
$H_\xi=\phi_0/(2\pi\xi^2)$.)

Fig. 2. ($a$) -- The maximal value of the order parameter, $\psi_{max}$, in
$m$-state; ($b$) -- the magnetization ($-4\pi M_\lambda$), Eq. (6);
$M_\lambda=M/H_\lambda$, $H_\lambda=\phi_0/(2\pi\lambda^2)$; ($c$) -- the
normalized free-energy, Eq. (7), as functions of $h_\xi$ for $R_\lambda=2$,
$\kappa=2$.

Fig. 3. The same, as in Fig. 2, but for: ($a,b,c$) -- $R_\lambda=2.5$,
$\kappa=1$; ($d,e,f$) -- $R_\lambda=4$, $\kappa=0.5$. The letters
$\alpha, \beta, \gamma, \delta, \varepsilon$ mark the points, where $m$-solution
terminates; the transitions  to the nearest laying energy states are shown by
the vertical arrows. The paramagnetic parts of the curves lay to the left of
vertical lines, where $M_\lambda=0$. [The arrows in Fig. 3($a$) show the
transition to the normal state ($\psi=0$). The arrows in Figs. 3($d$) mark the
possible hysteresis loop boundaries for the state $m=0$: $\gamma$ -- superheated
($sh$) Meissner state, $\delta$ -- supercooled ($sc$) normal state.]

Fig. 4. The co-ordinate dependences: ($a,d,g)$ -- the order parameter $\psi$,
($b,e,h$) -- the magnetic field $b$, ($c,f,i$) -- the current $j$; for solutions
with different $R_\lambda, \kappa, m$ (see numbers in the insets). The curves
{\it 1} correspond to maximal paramagnetism ($M>0$), the curves {\it 2}
correspond to $M=0$, the curves {\it 3} correspond to maximal diamagnetism
($M<0$). The curves {\it 1,2,3} are calculated for the following values of the
applied field $h_\lambda=H/H_\lambda$ ($h_\lambda=\kappa^2 h_\xi$):
{\it 1} -- $h_\lambda=1.52$,
{\it 2} -- $h_\lambda=2.921$,
{\it 3} -- $h_\lambda=4.30$ (if $R_\lambda=2, \kappa=2, m=4$);
{\it 1} -- $h_\lambda=0.45$,
{\it 2} -- $h_\lambda=0.985$,
{\it 3} -- $h_\lambda=1.52$ (if $R_\lambda=2.5, \kappa=1, m=2$);
{\it 1} -- $h_\lambda=0$,
{\it 2} -- $h_\lambda=0.195$,
{\it 3} -- $h_\lambda=0.48$ (if $R_\lambda=4, \kappa=0.5, m=1$).

Fig. 5. Shown are as functions of temperature $t=T/T_c$:
($a$) -- the maximal value of the order parameter, $\psi_{max}$;
($b$) -- the magnetization ($-4\pi M_\lambda$);
($c$) -- the normalized free-energy difference, $\Delta g$.
[For $m=0,1,2,3,4$ and $A=3$ ($\kappa=1$, $R=5.4\cdot 10^{-5}$\,cm,
$H=100$\,Oe), see Fig. 1($b$) and Eq. (9).]

\end{document}